\documentclass{vldb12}
\usepackage{graphicx}
\usepackage{balance}
\usepackage{color}
\usepackage[dvipsnames]{xcolor}
\usepackage{xspace}
\usepackage{textcomp}
\usepackage{amsmath}
\usepackage{hyperref}
\usepackage{balance}
\usepackage{caption, subcaption}
\usepackage[subtle]{savetrees}

\DeclareRobustCommand{\thetool}{{Foresight}\xspace}

\begin{document}
\title{\thetool: Recommending Visual Insights}
\numberofauthors{1}

\author {
  \c{C}a\u{g}atay Demiralp, Peter J. Haas, Srinivasan Parthasarathy, Tejaswini Pedapati \\
  IBM Research \\
  \{cagatay.demiralp, phaas, spartha, tejaswinip\}@us.ibm.com
}

\maketitle

\begin{abstract}

Current tools for exploratory data analysis (EDA) require users to manually
select data attributes, statistical computations and visual encodings. This can
be daunting for large-scale, complex data. We introduce \thetool,  a system
that helps the user rapidly discover visual insights from large
high-dimensional datasets. Formally, an ``insight'' is a strong manifestation of
a statistical property of the data, e.g., high correlation between two
attributes, high skewness or concentration about the mean of a single
attribute, a strong clustering of values, and so on. For each insight type,
\thetool initially presents visualizations of the top $k$ instances in the
data, based on an appropriate ranking metric. The user can then look at
``nearby'' insights by issuing ``insight queries'' containing constraints on
insight strengths and data attributes. Thus the user can directly explore the
space of insights, rather than the space of data dimensions and visual
encodings as in other visual recommender systems. \thetool also provides
``global" views of insight space to help orient the user and ensure a thorough
exploration process. Furthermore, \thetool facilitates interactive exploration 
of large datasets through fast, approximate sketching.

\end{abstract}

\section{Introduction}
\label{sec:intro}

Exploratory data analysis (EDA) is a fundamental approach for understanding and
reasoning about a dataset in which analysts essentially run mental experiments,
asking questions and (re)forming and testing hypotheses. To this end, analysts
derive insights from the data by iteratively computing and visualizing
correlations, outliers, empirical distribution and density functions, clusters,
and so on.

\noindent\textbf{EDA Challenges:} Although the capabilities of EDA tools continue to
improve, most tools often require the user to manually select among data
attributes, decide which statistical computations to apply, and specify
mappings between visual encoding variables and either the raw data or the
computational summaries. This task can be daunting for large datasets having
millions of data items and hundreds or thousands of data attributes per item,
especially for typical users who have limited time and limited skills in
statistics and data visualization. Even experienced analysts face cognitive
barriers in this setting. As discussed in~\cite{Pirolli_2005,Tversky_1975}, 
limitations on our working memory can cause large complex data to be
overwhelming regardless of expertise, and our tendency to fit evidence to
existing expectations and schemas of thought make it hard to explore insights
in an unbiased and rigorous manner. Thus people typically fail both to focus on
the most pertinent evidence and to attend sufficiently to the disconfirmation
of hypotheses~\cite{nickerson1998confirmation}.
Time pressures and data overload work against the analyst's ability to
rigorously follow effective methods for generating, managing, and evaluating
hypotheses.

\noindent\textbf{\thetool:} We attack this problem by introducing \emph{\thetool}, a
system that facilitates rapid discovery of visual insights from large,
high-dimensional datasets. \thetool enables users to jump-start the
exploration process from automatically recommended visualizations, and then
gives them increasing control over the exploration process as familiarity
with the data increases.  The resulting efficiency in insight generation can
save users time and dramatically improve their productivity, thereby expanding
the depth and breadth of generated hypotheses. Our approach, in which insights
are recommended according to objective criteria, also helps the analyst focus
more attention on evidence that is highly diagnostic for, or disconfirming to,
current hypotheses.

\noindent\textbf{Exploring Insight Space:} The key idea is to focus directly on
exploring the space of \textit{insights} rather than the usual space of data
dimensions and visual encodings, as in recent visualization recommendation
systems (e.g.,~\cite{Siddiqui_2016,Wongsuphasawat_2016}). We build on ideas
from prior research and commercial systems on automated and intelligent
analytics (e.g.,~\cite{WatsonAnalytics,MSPowerBI,Wills_2008}). Examples of
insights include a high linear correlation between attributes $x$ and $y$,
high concentration about the mean of $x$-values, the presence of extreme
$x$-value outliers, a strong clustering of $(x,y)$-values according to $z$-values, and so on.
Associated with each \textit{class} of insight are one or more strength metrics
that allow ranking---e.g., the Pearson correlation coefficient to measure the
strength of a linear correlation---as well as one or more visualization
methods. As discussed in Section~\ref{sec:model}, the metrics impose a
structure on insights that can be leveraged for exploration via \textit{insight
queries}. Given an unfamiliar, complex dataset, the user can select one or more
preliminary insights to investigate; in this first, open-ended stage of
exploration, \thetool visualizes the strongest examples of each insight. Using
an iterative procedure, the user can dive deeper into an insight class during a
second level of exploration by adding constraints on the data attributes
considered or on the values of the strength metric. Finally, each insight can
optionally support a third level of exploration by providing an
\textit{overview} visualization to help orient the user and ensure that the
exploration process is thorough.

\noindent\textbf{Sketching:} We use approximation techniques to achieve
interactive performance for insight queries. Specifically, the dataset is preprocessed to
compute sketches, samples, and indexes that will support fast approximate
insight querying. Importantly, we exploit the composability of certain types of
sketches to answer a broad range of insight queries.

Overall, \thetool contributes (i) a novel framework of insights, insight
metrics, insight visualizations and insight classes, (ii) sketch composition
for fast approximate computation of insight metrics and visualizations, and
(iii) an exploration engine for recommending and selecting insights that
satisfy user-specified constraints on strength and data attributes.

\section{Insights}
\label{sec:model}

\begin{figure*}[t]
  \includegraphics[width=\linewidth]{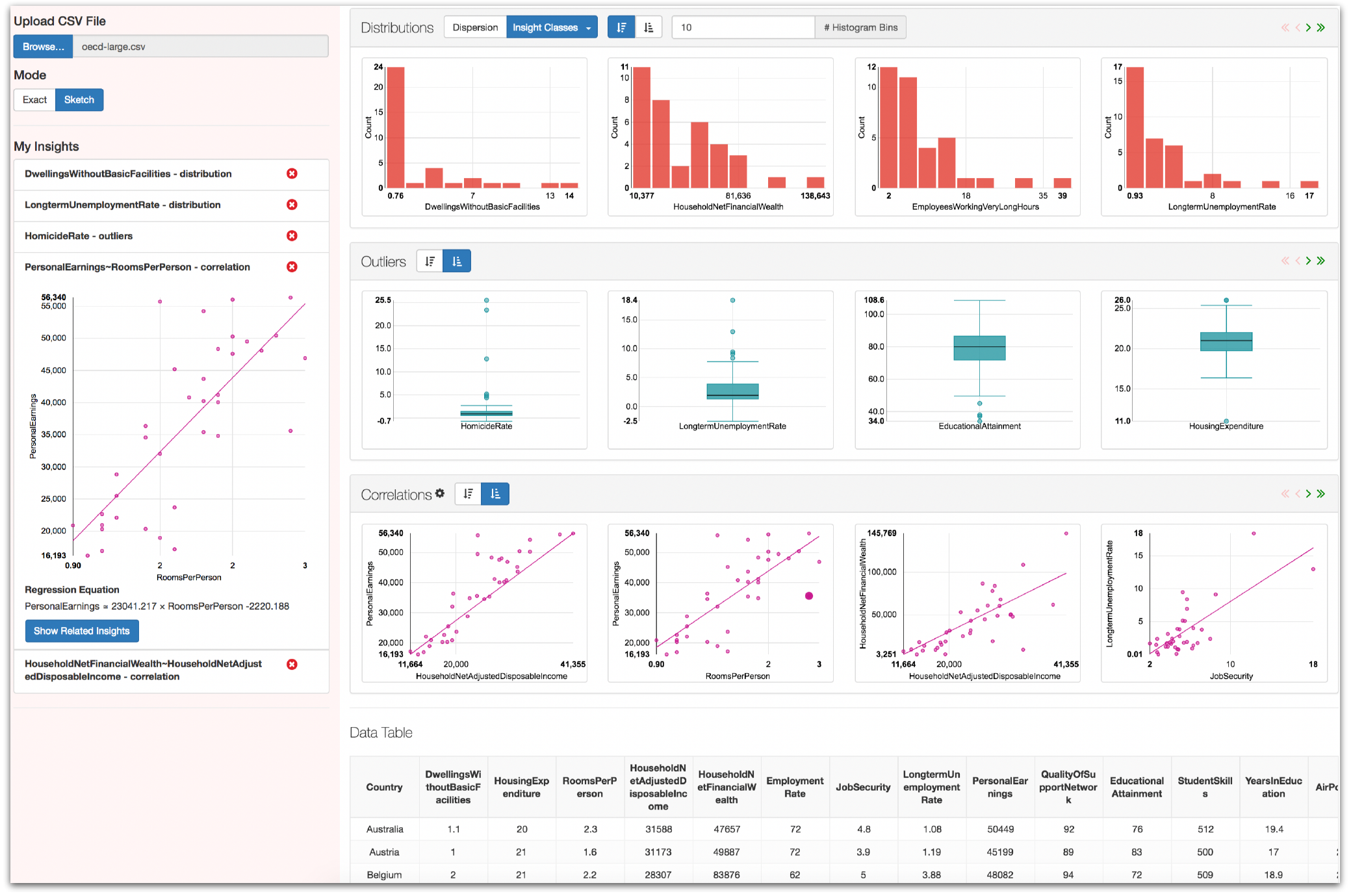}
  \caption{Each carousel in the Foresight UI corresponds to a distinct class of
    insight. Visualizations within a carousel are ranked by the insight's
    ranking metric with the strongest insights displayed first. In this
    screenshot, we show 3 of the 12 insight classes supported by Foresight,
    namely correlations, outliers, and heavy tails.\label{fig:insights}}
\end{figure*}

\begin{figure}[th]
  {\centering
    \includegraphics[width=\linewidth]{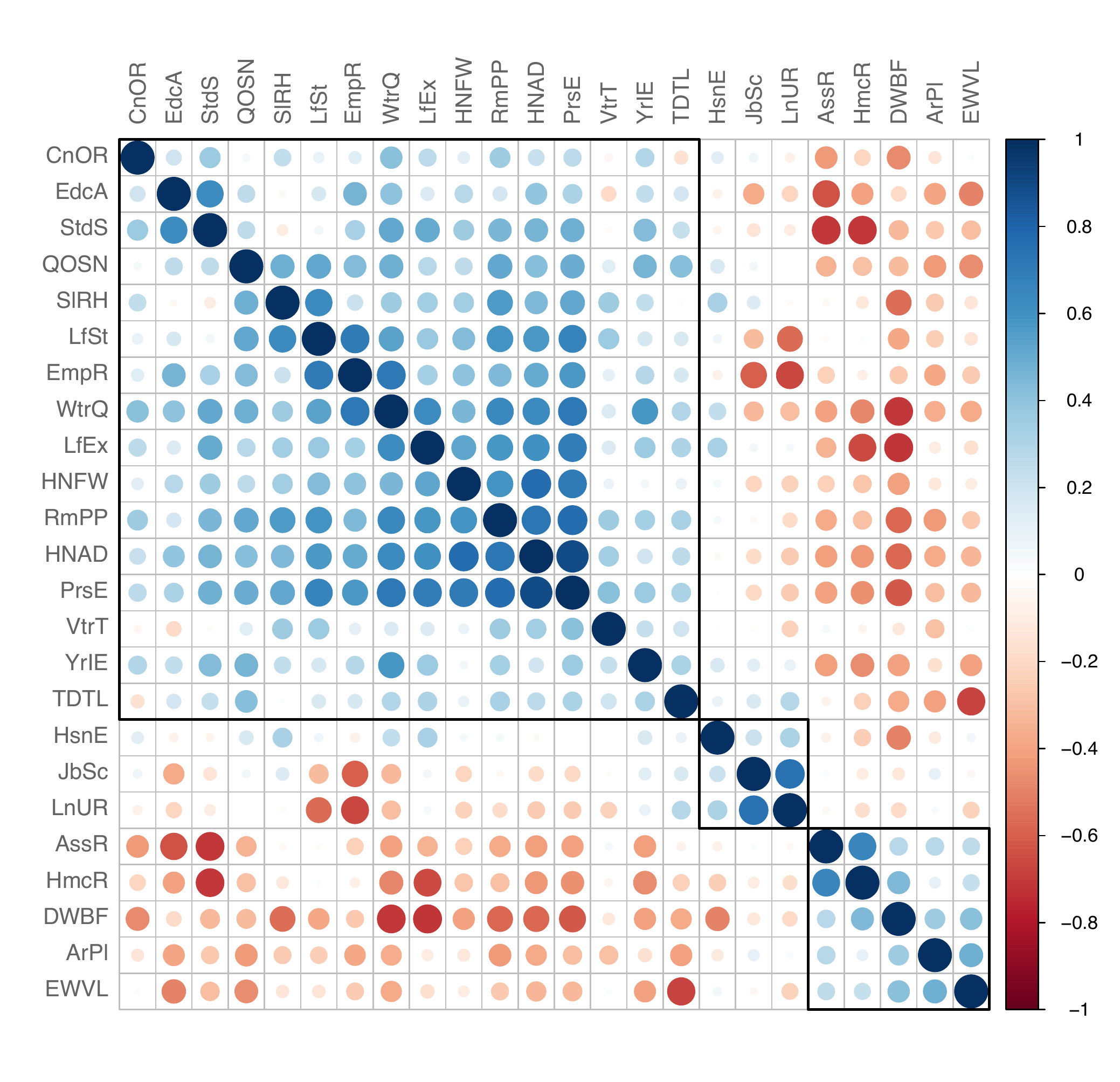}
    \caption{Each insight class can have an optional overview visualization
      of all the insights in the class. This overview displays all the
      pairwise attribute correlations as a heatmap with the size and
      intensity of circles denoting the strength of correlations.\label{fig:overview}}
   }
  \end{figure}

Here we first describe the basic concepts of queries on insight space
and then introduce the specific insights used in \thetool.

\subsection{Querying Insight Space}

The input data to \thetool is a matrix $A_{n \times d}$, where
each row represents one of $n$ \emph{data items} and each
column represents one of the $d$ \emph{attributes} of an item.
In this work, we assume the data has been pre-cleaned. In
general, the insights provided by \thetool might reveal
additional, more subtle data problems that require further
cleaning, e.g., a strong correlation that makes no real-world
sense.

We define an \emph{insight} as a strong
manifestation of a distributional property of the data, such as
strong correlation, tight clustering, low dispersion, and so
on. We focus throughout on insights involving the marginal
distribution of one, two, or three attributes (Figure \ref{fig:insights}).

We require that each insight have one or more associated
\emph{insight metrics} that can be used to rank $n$-tuples of
attributes based on the strength of the property that
defines the insight. Similarly, each insight must have one or
more associated \emph{data visualizations}. Corresponding to
each insight is an \emph{insight class} that comprises all
feature tuples whose joint distributions are compatible with
the insight's associated metrics and visualizations. For
example, given a data set with attributes $a_1,a_2,\ldots,a_n$,
the insight class corresponding to the insight $\mathcal{I}=$
``high linear correlation" would contain all pairs $(a_i,a_j)$
with $i<j$ such that $a_i$ and $a_j$ are both real-valued
attributes. Finally, an insight may optionally have one or more
associated \emph{overview visualizations} that display the
values of the insight metric over all tuples in the insight
class. The global visualization for the insight~$\mathcal{I}$
above, for example, is a heat map where the $x$ and $y$
coordinates correspond to the different attribute indices and
the color and size of the circle centered at $(x,y)$ encode
the Pearson correlation coefficient (Figure \ref{fig:overview}).

A basic \emph{insight query} returns the visualizations for the
highest-ranked feature tuples according to the insight
metric selected, e.g., the attribute pairs with the highest
correlations. In general, one or more of the attributes might
be fixed, e.g., instead of ranking the highest correlations
over all $(x,y)$ attribute pairs, we can fix $x=\bar{x}$ and
rank correlations only over pairs of the form $(\bar{x},y)$,
i.e., searching for the attributes most correlated with $\bar{x}$.
Insight queries may also have constraints or filters on the
strength metric, e.g., we might want to rank only correlated
attribute pairs whose correlation coefficient falls in the
range $[0.5,0.8]$ because we want to filter out trivially very
high correlations. In future work, queries will also allow
inclusion of constraints involving metadata about attributes,
e.g., to search for attributes that represent currency or
dates.

As can be seen, our framework imposes some structure on the
space of insights that can be exploited during search. Two
insights can be considered ``similar" if their metric scores
are similar or if the sets of fixed attributes are similar. At
any point during the EDA process, the user can step back and
look at the overview visualization of an insight (Figure~\ref{fig:overview}). This helps
ensure that, in analogy with gradient descent,  the EDA process
does not get inadvertently ``trapped" in some local
``neighborhood" of  attribute tuples. This capability is
particularly important in cases where many attribute tuples
have similarly high insight-metric scores, so that the
particular set visualized for the user is somewhat arbitrary.
Section~\ref{sec:demo} illustrates the insight-navigation
process in a concrete scenario.

\subsection{\thetool's Insight Classes}

Foresight is designed to be an extensible system where a data scientist
can ``plug in" new insight classes along with their corresponding
ranking measures and visualizations. We now briefly describe some of the specific insights supported by
\thetool. Denote by $\mathcal{B}$ and $\mathcal{C}$ the sets of
attribute columns in $A$ that contain numeric and categorical
values. \thetool supports a variety of distinct visual insights, each
with a preferred ranking metric and visualization method. Denote by
$b=(b_1,\ldots,b_n)^\top\in\mathcal{B}$ a numeric column with
mean $\mu_b$ and standard deviation $\sigma_b$, and by
$c=(c_1,\ldots,c_n)^\top\in\mathcal{C}$ a categorical column.
For each insight, the \textit{ranking metric} is italicized.

\noindent\textbf{1. Dispersion:} Very high or low dispersion of
data values around a population mean is measured by the
\textit{variance} $\sigma^2(b) = n^{-1}\sum_i^n (b_i - \mu_b)$
and is visualized via a histogram.

\noindent\textbf{2. Skew:} Skewness is a measure of asymmetry
in a univariate distribution. It is measured by the
\textit{standardized skewness coefficient} $\gamma_1(b) =
n^{-1}\sum_i^n (b_i - \mu_b)^3/\sigma_b^3$ and visualized via a
histogram.

\noindent\textbf{3. Heavy Tails:} Heavy-tailedness is the
propensity of a distribution towards extreme values. It is
measured by \textit{kurtosis} $\text{Kurt}(b) =n^{-1} \sum_i^n
(b_i - \mu_b)^4/\sigma_b^4$ and visualized via a histogram.

\noindent\textbf{4. Outliers:} The presence and significance of
extreme outliers is measured by applying a user-configurable
outlier-detection algorithm---see, e.g.,
\cite{Aggarwal13}---and computing the \textit{average
standardized distance} of the outliers from the mean, where
standardized distance is measured in standard deviations.
Outliers are visualized using box-and-whisker plots.

\noindent\textbf{5. Heterogeneous Frequencies:} For a
categorical column $c$ (or a discrete numerical column $b$),
high heterogeneity in frequencies implies that a few values
(``heavy hitters'') are highly frequent while others are not.
For a configurable parameter~$k$, heterogeneity strength is
measured by $\text{RelFreq}(k,c)$, the \textit{total relative
  frequency of the~$k$ most frequent elements in $c$}. This
  insight is visualized via a Pareto chart.

\noindent\textbf{6. Linear Relationship:} The strength of a
linear relationship between two columns $x, y \in \mathcal{B}$
is measured using the magnitude of the \textit{Pearson correlation coefficient} $|\rho(x,y)|$, where 
$\rho(x, y) = \sum_{i=1}^n
(x_i-\mu_x)(y_i-\mu_y)/(\sigma_x\sigma_y)$ and visualized via a
scatter plot with the  best-fit line superimposed.

\noindent\textbf{Additional Insights:} Other insights include multimodality, nonlinear monotonic relationships, general statistical dependencies, and segmentation. Details are suppressed due to lack of space.

\section{Sketching}
\label{sec:sketch}

We use sketching~\cite{CormodeGHJ12} to speed up the computation of insight
metrics.  Some insight metrics are fast and easy to compute, e.g., skewness and
kurtosis can both be computed for numeric columns in a single pass by
maintaining and combining a few running sums. For the remaining metrics,
sketches---lossy compressed representations of the data---are crucial in order
to preprocess the data in a reasonable amount of time. \thetool integrates and 
composes a variety of sketching and sampling techniques from the
literature, namely quantile sketch, entropy sketch, frequent items sketch,
random hyperplane sketch, and random projection sketch; see, e.g.,
\cite{CormodeGHJ12}.
As an illustration, we describe the use of the random hyperplane
sketch~\cite{moses:random} in \thetool for approximating a Pearson 
correlation coefficient $\rho$.

To create the sketch, we first generate $k$ distinct random vectors $r_1,
\ldots, r_k$, where $k\ll n$ and each $r_i$ is $n$-dimensional with components
drawn independently from the one-dimensional standard normal distribution. For
each $r_i$, define a function $\phi_i$ by
\[
\phi_i(b) =
\begin{cases}
0 & \text{if $\tilde{b} \cdot r_i < 0$};\\
1 & \text{if $\tilde{b}  \cdot r_i \ge 0$}
\end{cases}
\]
for $b\in\mathcal{B}$, where $\tilde{x}$ denotes the ``centered" version of
column $x$ obtained by subtracting $\mu_x$ from each component. Then the sketch
for a specific column $b$ is the random bit-vector
$\bigl(\phi_1(b),\ldots,\phi_k(b)\bigr)$, which we write as $\phi(b)$.  For
$n$-dimensional vectors $x,y\in\mathcal{B}$, set $\delta_j=1$ if $x_j\not=y_j$
and $\delta_j=0$ otherwise ($1\le j\le n$), and define the \emph{Hamming distance}
between $x$ and $y$ by $H(x,y)=\sum_{j=1}^n \delta_j$. As shown
in~\cite{moses:random}, the quantity $\cos(\pi H_{xy}/k)$, where $H_{xy}=
H\bigl(\phi(x),\phi(y)\bigr)$, is an unbiased estimator of the correlation
coefficient $\rho(x,y)$.

The bit-vector sketch consumes $|\mathcal{B}|k$ bits of memory for the entire
dataset and can be computed in a single pass of the data in time
$O(|\mathcal{B}|nk)$. Furthermore, computing the estimated correlation
coefficient between every pair of features takes $O(|\mathcal{B}|^2k)$ time as
opposed to $O(|\mathcal{B}|^2n)$ time. Setting $k$ to a value that is
$O(\log^2{n})$ guarantees high accuracy while significantly reducing the time 
complexity of ranking and searching for correlation coefficients.  

Initial experiments (without parallelism) showed
$>90\%$ accuracy and $3x-4x$ speedup in preprocessing, with interactive speeds
during exploration.

\section{Demonstration}
\label{sec:demo}
  \subsection{Usage Scenario}
  We now describe how an analyst uses \thetool to explore a dataset
  containing wellbeing indicators for the OECD member countries. This dataset
  contains 25 distinct attributes (indicators) about 35 countries and is included
  in our demo as an illustration and for ease of comprehension. \thetool
  is intended to facilitate interactive exploration of datasets with data items
  of the order of 100K and attributes that number in the hundreds.

  The analyst loads the OECD dataset in Foresight and eyeballs various insights
  displayed in the carousels corresponding to each insight class (Figure
  \ref{fig:insights}).  She notes instantly that the indicators {\tt
  Working Long Hours} and {\tt Time Devoted To Leisure} have a strong negative
  correlation, since this is one of the top-ranked correlation insights
  recommended by Foresight. Encouraged by this quick discovery, she brings
  this insight \textit{into focus} by clicking on it.  \thetool updates its
  recommendations by choosing a subset of insights within the neighborhood of the
  focused insight. The analyst explores the newly recommended correlations
  through multiple ranking metrics such as Pearson correlation coefficient and
  Spearman rank correlation and is surprised to learn that {\tt Time Devoted To
  Leisure} has no correlation with {\tt Self Reported Health}.

  Intrigued with this lack of correlation, she checks the univariate
  distributional insight classes. The recommendations within these classes,
  which have already been updated based on the previous selections, show that
  {\tt Time Devoted To Leisure} has a Normal distribution while {\tt Self Reported
  Health} has a left-skewed distribution. Having gained greater familiarity
  with the OECD dataset, our analyst wonders about the factors that affect
  {\tt Self Reported Health}. She clicks on the distribution of \texttt{Self
  Reported Health}, adding this as one of the focal insights. Foresight
  recommends a new set of correlated attributes and she finds that {\tt
  Life Satisfaction} and {\tt Self Reported Health} are highly correlated.

  Satisfied with her preliminary discoveries (and armed with deeper questions
  about OECD countries than before), our analyst saves the current
  \thetool state to revisit later and to share with her colleagues.

  \balance
  \subsection{Demo Datasets}
  Our demonstration will feature the following two datasets in addition to the
  OECD dataset described above.

  \noindent\textbf{Parkinson:} Parkinson's Disease (PD) is a progressive
  neurodegenerative disorder affecting nearly a million people in
  the US alone. Our second use case applies \thetool to gain insights
  into a dataset of PD patients with measured clinical descriptors
  characterizing the disease progression~\cite{goetz2008movement}. The dataset
  has 2K rows and 50 columns and is collected under the Parkinson's
  Progression Markers Initiative (PPMI).

  \noindent\textbf{IMBD:} Our third use case explores a dataset with 5000
  movies (rows) and 28 features (columns). The features range from
  the director name to the IMBD score for each movie. Questions that
  Foresight users will be able to explore are: What factors correlate
  highly with a film's profitability?  How are critical responses and
  commercial success interrelated?





\section{Conclusion}
\label{sec:conclusions}

We introduce a novel approach to visualization recommendation via the notion of
insights. Our approach uses recommendations to guide users in exploring
unfamiliar large and complex datasets, and gradually gives them more and more
control over the exploration process. Using sketching and indexing methods, our
demo system can currently handle datasets with large numbers of rows and
moderate numbers of columns. Future work will improve the scalability with
respect to columns by incorporating parallel search methods that speed up
insight queries.

\bibliographystyle{abbrv}
\bibliography{refs}

\end{document}